\documentstyle[aps,multicol]{revtex}

\input{tcilatex}

\begin{document}
\title{Observation of spin freezing and relaxation at microwave frequencies in the
spin ladder compound $Sr_{14-x}Ca_{x}Cu_{24}O_{41}$ }
\author{Z. Zhai, P. V. Patanjali, N. Hakim, J. B. Sokoloff and S. Sridhar$^{a}$}
\address{Physics Department, Northeastern University, 360 Huntington Avenue, Boston,\\
MA 02115}
\author{U. Ammerahl, A. Vietkine and A. Revcolevschi}
\address{Laboratoire de Chimie des Solides, Universite Paris-Sud, Orsay, France}
\maketitle

\begin{abstract}
We report the observation of a frequency ($\omega $) and temperature ($T$)
-dependent loss peak in $\chi ^{\prime \prime }$ and accompanying dispersion
in $\chi ^{\prime }$ from microwave ($2-18GHz$) measurements of the complex
susceptibility $\tilde{\chi}(\omega ,T)=\chi ^{\prime }+i\chi ^{\prime
\prime }$ of $Sr_{14-x}Ca_{x}Cu_{24}O_{41}$. We associate this phenomenon
with a rapid decrease of spin disorder and corresponding spin relaxation
rate, representing a ``spin freezing transition'' accompanying charge
ordering which occurs at temperatures $\sim 250K$ (for the $x=0$ compound).
Our results enable direct quantitative measurements of the spin relaxation
rate, and yield information on spin dynamics in these materials and the
related compounds $SrCuO_{2}$ and $Sr_{2}CuO_{3}$.
\end{abstract}

\begin{multicols}{2}%

The dynamics of the basic $Cu-O$ building blocks of the high temperature
superconducting cuprates is of great interest, as they may provide insights
into the fundamental magnetic and superconducting interactions in these
materials. To this end the synthesis of the structurally simpler but related
chain / ladder compounds in the $Sr-Cu-O$ family represented an important
step. It has since become clear that these spin chain / ladder compounds
possess unique and interesting quantum magnetic properties \cite{dagotto93}
and also display superconductivity \cite{Uehara96}.

\smallskip In this paper, we describe some novel observations related to
spin dynamics at microwave frequencies in the spin chain / ladder compounds $%
Sr_{14-x}Ca_{x}Cu_{24}O_{41}$, $SrCuO_{2}$ and $Sr_{2}CuO_{3}$. In precision
measurements at microwave frequencies $(2-18GHz)$ of the dynamic magnetic
susceptibility $\tilde{\chi}(\omega ,T)=\chi ^{\prime }+i\chi ^{\prime
\prime }$ of $Sr_{14}Cu_{24}O_{41}$, we observe a microwave loss peak
appearing as a maximum in the microwave absorption (i.e. $\chi ^{\prime
\prime }$), and accompanied by a rapid decrease of the dispersion $\chi
^{\prime }(\omega ,T)$, both as functions of temperature. These features are
strongly frequency dependent in the microwave frequency range and are not
observed in static (dc) magnetization or resistivity measurements. The cause
of this is a rapid decrease of the spin-spin relaxation rate which is
extracted from the data, and indicates a rapid decrease in spin disorder
below approximately $250K$ in the $x=0$ compound. The peak temperature
decreases with $Ca$ doping. This is a dynamic magnetic signature of the
onset of charge ordering suggested by structural \cite{Cox98} and NMR
measurements \cite{Takigawa98}. Additional magnetic relaxation processes due
to holes at high temperatures and dimer formation at low temperatures are
also observed. The precision microwave measurements, which have a dynamic
range over several orders of magnitude, are thus a sensitive probe of
electron spin relaxation in condensed matter systems.

Samples of $Sr_{14}Cu_{24}O_{41}$, $Sr_{14-x}Ca_{x}Cu_{24}O_{41}$($%
x=2,8,11.5 $), $SrCuO_{2}$ and $Sr_{2}CuO_{3}$ were prepared by the floating
zone technique \cite{Vietkine}. These single crystal samples have been
extensively characterized by a vast array of measurements : dc resistivity,
ac and dc SQUID susceptibility, XRD, neutron scattering, high pressure
studies, and numerous other techniques. All of these measurements are in
good agreement with those reported in the literature and indicate single
phase, high quality crystals.

The principal measurements reported here are carried out using a
superconducting microwave cavity at $10GHz$, with additional measurements at 
$2GHz$ and $18GHz$. These experiments have been extensively utilized
previously for measuring a variety of materials, including superconducting
cuprate, nickelate and borocarbide crystals \cite{Sridhar88,Srikanth98}.
Typical samples were $2\times 2\times 2$ $mm^{3}$ in size. The anisotropic
response was studied by orienting the sample with the microwave magnetic
field $H_{\omega }$ parallel to one of the $a,b,c$ crystal axes. In all of
the measurements, the sample was placed in maximum microwave magnetic field
and in zero microwave electric field. The cavity parameters can be related
to the sample microwave susceptibility $\tilde{\chi}$ by : $%
f(0)-f(T)+i\Delta f=a(\delta \chi ^{\prime }(T)+i\chi ^{\prime \prime }(T))$%
, where $f(T)$ is the resonant frequency, $\Delta f(T)$ is the width of the
resonance, and $a$ is a sample geometric factor. While the loss term $\chi
^{\prime \prime }(T)$ is measured absolutely, the technique yields changes $%
\delta \chi ^{\prime }(T)=\chi ^{\prime }(T)-\chi ^{\prime }(0)$ in
susceptibility with very high precision.

The microwave susceptibility $\delta \chi _{c}^{\prime }(10GHz,T)$ and $\chi
_{c}^{\prime \prime }(10GHz,T)$ versus $T$ for $Sr_{14}Cu_{24}O_{41}$ are
shown in Fig.\ref{fig1}. The subscript $c$ indicates that the microwave
field $H_{\omega }\,||\,c$, i.e. parallel to the chains. Similar results
were obtained on measurements of other crystal samples of this material. The
most striking feature of the data is the rapid drop with decreasing $T$ in
susceptibility $\delta \chi _{c}^{\prime }(10GHz,T)$ below approximately $%
200K$. This is accompanied by a relatively sharp peak in $\chi _{c}^{\prime
\prime }(10GHz,T)$, with a peak temperature of $T_{p10GHz}\sim 170K$. It is
worth noting that below the peak the absorption has dropped by nearly a
factor of $100$ between $170K$ and $100K$. Indeed the rapid decreases in
susceptibility ($\delta \chi _{c}^{\prime })$ and large absorption($\chi
_{c}^{\prime \prime }$) can be easily mistaken for a (broad) superconducting
transition.

Another striking aspect is a very strong frequency dependence. At $2GHz$,
the peak is shifted downward to $T_{p2GHz}\sim 152K$ while at $18GHz$ it is
shifted upwards to $T_{p18GHz}\sim 189K$. The strong variation with
frequency of the temperature dependence of $\delta \chi _{c}^{\prime }(f,T)$
and $\chi _{c}^{\prime \prime }(f,T)$ is shown in Fig.\ref{fig2}. Due to
differences in the techniques, the measurements at different frequencies are
shown in arbitrary units.

It is important to note that {\em similar features are not observed in
static and low frequency measurements}. dc SQUID susceptibility measurements
shown in Fig.\ref{fig1} (inset, bottom panel) are consistent with those
available in the literature \cite{Carter96} and only indicate a gradual
increase of the susceptibility with decreasing $T$. We have also measured
the susceptibility at $1KHz$ and $3MHz$ and do not see these features. Thus
the phenomenon seen in Fig.\ref{fig1} is observable only at high frequencies
in the microwave spectral range.

We have carried out extensive measurements to check the validity of these
remarkable results. Similar results were observed in 3 samples. In
particular we have confirmed that these data cannot arise from a purely
resistive (eddy-current) sample size effect, since the resistivity increases
with decreasing $T$, while the observed {\em \ }$\delta \chi _{c}^{\prime
}(10GHz,T)$ in Fig.\ref{fig1} can only be consistent with $\rho (T)$
decreasing with decreasing $T$. (The eddy-current contribution is relevant
at high temperatures and is included in the analysis later). Taken together
these checks confirm that the observed phenomenon is magnetic and due to the
spin system.

Important insights can be gained by plotting the data as $\chi _{c}^{\prime
\prime }$ vs. $\delta \chi _{c}^{\prime }$, as shown in Fig.\ref{fig3}. This
so-called ``arc-plot'' can be regarded as the magnetic equivalent of a
Cole-Cole plot used in dielectric spectroscopic studies of liquids \cite
{Wei90}. However here we are varying temperature as a parameter rather than
frequency. The data show a dominant relaxation process represented by the
nearly semi-circular arc for most of the temperature $100K$ to $220K$, above
which temperatures another process dominate.

A close examination shows that the totality of our results can be expressed
as $\tilde{\chi}_{c}(\omega ,T)=\tilde{\chi}_{c\alpha }(\omega ,T)+\tilde{%
\chi}_{c\beta }(\omega ,T)+i$ $\chi _{c\gamma }^{\prime \prime }(T)$, where $%
\tilde{\chi}_{c\alpha }(<<\tilde{\chi}_{c\beta })$ represents a low
temperature dimer contribution which is discussed later, $\tilde{\chi}%
_{c\beta }$ the loss peak and accompanying susceptibility transition in
Fig.1 and Fig. 3, and $\tilde{\chi}_{c\gamma }$ the high temperature
relaxation which we attribute to the eddy-current contribution arising from
the finite resistivity. (See Fig.\ref{fig1} top panel for the location in $T$
of these processes).

The eddy-current term due to the resistivity $\rho (T)$ can be written as $%
\tilde{\chi}_{c\gamma }(T)=\tilde{g}(2a/\delta )$ \cite{Hein91}, where $%
\tilde{g}$ is a function of the ratio of the sample radius $a$ and the skin
depth $\delta =(2\rho /\omega \mu _{0})^{1/2}$. While the exact expressions
depend on the sample shape, we have carried out careful numerical analysis
and verified that the present data are in the limit $a<<\delta $, where $%
\chi _{c\gamma }^{\prime \prime }\approx (2a/\delta )^{2}/20$. From the
experimental data, we get $\chi _{c\gamma }^{\prime \prime }\sim $ $1.5\exp
(-\Delta _{c\gamma }/kT)$, which is consistent with an activated resistivity 
$\rho (T)$ with a gap value of $\Delta _{c\gamma }=1200K$, and is hence in
reasonable agreement with typical dc resistivity measurements \cite{Carter96}%
. Note that in this regime $\chi _{c\gamma }^{\prime }<<\chi _{c\gamma
}^{\prime \prime }$. The numerical estimates show that the $\chi _{c\gamma }$
process leads to a peak in $\chi _{c\gamma }^{\prime \prime }$ at much
higher temperatures $>>300K$, accompanied by a corresponding {\em decrease}
in $\chi _{c\gamma }^{\prime }$ the same temperature.

Subtracting the $\chi _{c\gamma }^{\prime \prime }=1.5\exp (-1200/T)$
contribution leads to a nearly pure relaxation $\tilde{\chi}_{c\beta }$ as
is evident from the arc plot in Fig.\ref{fig3}(bottom curve). We therefore
analyze the results in terms of spin relaxation leading to a susceptibility $%
\tilde{\chi}_{c\beta }(\omega ,T)=\Delta \chi _{\beta }/(1+i\omega \tau
_{\beta }(T))$. (Although the data of Fig.3 do not lie exactly on a
semi-circle, they are close enough to justify the use of a single relaxation
time for convenience). As the relaxation time $\tau (T)$ changes with $T$, a
loss peak occurs in the absorption at the temperature $T_{p}$ when $\omega
\tau (T_{p})=1$, i.e. when the relaxation rate $(2\pi \tau (T_{p}))^{-1}$%
crosses the measurement frequency $f$. This is accompanied by a net change
of the susceptibility $\Delta \chi _{\beta }$. The sharpness of the peak is
determined by the rate of variation of $\tau _{\beta }$ with $T$.(Note that
the peak in the measured $\chi _{c}^{\prime \prime }$ is at a different
location than that inferred from $\chi _{c\beta }^{\prime \prime }$ because
of the contribution of the $\gamma $ relaxation).

The relaxation time $\tau _{\beta }(T)$ can be directly obtained from the
data of Fig.\ref{fig1}, as $\tau _{\beta }(T)=\omega ^{-1}(\chi _{c_{\beta
}}^{\prime \prime }(T)/\chi _{c\beta }^{\prime }(T))$. This is shown in Fig.%
\ref{fig4} as the relaxation rate $(2\pi \tau _{\beta })^{-1}(T)$. The
relaxation rate $\tau _{\beta }^{-1}(T)$ clearly drops dramatically in the
vicinity of approximately $250K$. The relaxation rate is due to spin-spin
interactions and can therefore be related to a spin fluctuation rms field $%
H_{i}$ from the relation $\tau =h/(8\pi g\mu _{B}H_{i})$. Thus the data in
Fig.\ref{fig4} indicate a rapid slowing of spin fluctuations below $250K$
leading to a sharp drop in the relaxation rate. The phenomenon represented
by Fig. 1 is very similar to the {\em ``spin freezing transition''} that is
seen in probes of spin dynamics such as NMR (e.g. ref.\cite{Hammel93}) and $%
\mu SR$ \cite{Niedemayer98} but characterized there by lower (MHz)
frequencies. A key feature of our present experiments is the much higher
frequency range and the direct coupling to the electron spins.

There are several other experiments \cite{Takigawa98,Imai98,Cox98} which
have observed anomalies in the vicinity of $200K$ in $Sr_{14}Cu_{24}O_{41}$.
NMR and NQR experiments \cite{Takigawa98} reported that the $Cu^{3+}$ NMR
signal splits into 2 peaks below $\sim 200K$, suggesting the occurrence of
charge ordering. Neutron scattering experiments \cite{Eccleston98} have
reported charge ordering at $50K$ eventually disappearing close to $300K$.
Cox et al.,\cite{Cox98} interpret their synchrotron X-ray scattering results
in terms of a charge-ordered model involving both dimerization between two
next-neighbor $Cu^{2+}$ ions surrounding a $Cu^{3+}$ ion on a Zhang-rice
singlet site comprised of a $Cu^{2+}$ ion and a hole on the surrounding $O$
atoms, and dimerization between $Cu^{2+}$ ions, in the linear $Cu-O$ chains.
Thus the NMR and synchrotron radiation experiments on $Sr_{14}Cu_{24}O_{41}$
are consistent with an ordered arrangement of $Cu^{2+}$ and Zhang-Rice
singlets below approximately $300K$. Such a spin ordering reduces the spin
susceptibility. {\em However, our data shows that there is no static
ordering, but only the a reduction in the rate of fluctuations of the spins.}

Further insight into the microscopic origin of this phenomenon can be
achieved by comparing with our measurements of $\delta \chi _{c}^{\prime
}(10GHz,T)$ and $\chi _{c}^{\prime \prime }(10GHz,T)$ versus $T$ for $%
SrCuO_{2}$. This is shown in Fig.\ref{fig5}. Although the magnitudes of the
susceptibilities are a factor $10^{2}$ lower in $SrCuO_{2}$, a weak peak is
also observed, which is however not as sharp, indicating that the spin
relaxation rate does not vary as strongly as in $Sr_{14}Cu_{24}O_{41}$,
probably because of the smaller hole density, and the absence of any
reported charge ordering. It is worth noting that static susceptibility
measurements on $SrCuO_{2}$ do not show any features at this temperature\cite
{Matsuda97,Motoyama96}. Since a structural unit common to both is the
zig-zag $CuO_{2}$ chain, the data suggests that this is the magnetic
sub-unit that is dominant at these temperatures in both $%
Sr_{14}Cu_{24}O_{41} $ and $SrCuO_{2}$.

In the zig-zag chains in $SrCuO_{2}$ both $J_{1}$ (the $180^{\circ }$ $%
Cu-O-Cu$ AFM interaction) and $J_{2}$ (the nearly $90^{\circ }$ FM
interaction) are present. In $Sr_{14}Cu_{24}O_{41}$ there is additional
coupling $(J_{3})$ along the ladder rung. Competition between $J_{1}$, $%
J_{2} $ and $J_{3}$ could lead to frustration. The increase in the strength
of the loss peak (Fig. 5) by orders of magnitude in $Sr_{14}Cu_{24}O_{41}$
could possibly be due to enhanced frustration, besides the increasing hole
density, both leading to more dissipation. It has, recently, been suggested
theoretically that the frustrated zigzag chain-ladder system possesses
incommensurate spin correlations \cite{Nersesyan98}. The present results
(Fig. 5) suggest that incommensurate spin correlations in the zig-zag chains
caused by frustration may also be responsible for the loss peak.

The above conclusion is further confirmed by measurements on $Sr_{2}CuO_{3}$
(Fig.\ref{fig5}) which only possesses linear $Cu-O$ chains and is regarded
as an ideal 1-D Heisenberg AFM. Here the microwave measurements show only
signatures of the 3-D static AFM order below $T_{N}=5.5K$\cite{Motoyama96}.
At high temperatures in the vicinity of $160-300K$ the features discussed
above in $Sr_{14}Cu_{24}O_{41}$and $SrCuO_{2}$ are absent.

We have studied the systematics of $\tilde{\chi}(\omega ,T)$ with $Ca$
doping (see Fig.\ref{fig5}). Focussing on the peak in $\chi _{c\beta
}^{\prime \prime }(10GHz,T)$ for the present, we find that it is shifted to
a lower temperature $140K$ in $Sr_{12}Ca_{2}Cu_{24}O_{41}$with greatly
decreased amplitude. It is further shifted to a very broad, weak peak at
around $110K$ in $Sr_{4}Ca_{8}Cu_{24}O_{41}$. In $%
Sr_{2.5}Ca_{11.5}Cu_{24}O_{41}$ it is not visible and is perhaps overwhelmed
by a low temperature ($\tilde{\chi}_{c\alpha }$) process. An extensive
discussion of these results of the doping dependence, as well as extensive
studies of anisotropy where the field $H_{\omega }||\,b$ and $a$ axes in $%
Sr_{14-x}Ca_{x}Cu_{24}O_{41}$ for $x=0,2,8,11.5$ will be provided in a
detailed publication.

The extraordinary sensitivity of the superconducting cavity measurements
enables us to see additional phenomena (which we call $\tilde{\chi}_{c\alpha
}$) at low temperatures. This is evident from the semi-log plot in Fig.\ref
{fig1}(bottom panel), which reveals a low temperature feature in $%
Sr_{14}Cu_{24}O_{41}$ below about $100K$. This temperature scale has been
identified with singlet dimer formation in the $CuO_{2\text{ }}$chains with
a spin gap value $140K$, as has been observed in ac susceptibility and
NMR/NQR studies \cite{Matsuda97}. As two neighboring $Cu^{2+}$ spins in the
1D chains couple antiferromagnetically to form a singlet, most of the $%
Cu^{2+}$ (S=1/2) spins among the comparable amount of $Cu^{3+}$ (S=0) ions
will form dimers\cite{Matsuda97}. This mechanism has been used to explain
the dc magnetic susceptibility \cite{Matsuda97} and is likely to be
responsible here also for the low temperature features in Fig.\ref{fig1},
modified by frequency-dependent corrections. Note also that this is the
first measurement of $\chi _{c\alpha }^{\prime \prime }$ and hence spin
relaxation at these temperatures.

The present results show clearly the presence of spin relaxation mechanisms
with time scales corresponding to the GHz frequency ranges. These results
have a greater relevance to superconductivity in the cuprates since the
presence of strongly $T-$dependent relaxation times is essential to
understanding the microwave response of the cuprate superconductors \cite
{Srikanth98}. The microwave measurements yield information on dynamics at
short time scales $10^{-11}$ sec. and longer, comparable to NS but shorter
than NMR and NQR ($10^{-7}$ sec.) and $\mu SR$ ($10^{-8}$ sec.). Thus the
present precision measurements are a new probe of spin dynamics in the form
of a spin relaxation spectroscopy in quantum magnets and superconductors.
This will also require further theoretical developments, such as microscopic
calculations of $\tilde{\chi}(\omega ,T)$, to quantitatively describe the
various interactions observed in the present work.

This research was supported by NSF-9623720 and by a US-France collaborative
grant NSF-INT-9726801. We thank R. S. Markiewic and C. Kusko for useful
discussions.

$^{a}$Electronic address: srinivas@neu.edu

\end{multicols}%

\begin{figure}
\caption{(Top) Microwave susceptibility $\delta \chi_c ^{\prime }(10GHz,T)$
and $\chi_c ^{\prime \prime }(10GHz,T)$ versus $T$ for $Sr_{14}Cu_{24}O_{41}$.
(Bottom) Semilog plot of the data in the top panel. The large dynamic range of the experiment is evident.
Note the large drop in absorption, i.e. $\chi_c^{\prime\prime}$. This plot also shows the 
low $T$ $\chi _{c \alpha}$ process below $100K$.
(Inset, bottom panel) dc susceptibility $\chi (10^{-3}$ $emu/mole-Cu)$  which shows no features between $100K$ and $300K$.}
\label{fig1}
\end{figure}%

\begin{figure}
\caption{ $\delta \chi_c ^{\prime }(f,T)$
and $\chi_c ^{\prime \prime }(f,T)$ versus $T$ for $Sr_{14}Cu_{24}O_{41}$, at $f$ = 2, 10 and 18
GHz. Note that the peak temperature is frequency dependent.}
\label{fig2}
\end{figure}%

\begin{figure}
\caption{$\chi_c ^{\prime \prime }(f,T)$ {\em vs.}  $\delta \chi_c ^{\prime }(f,T)$ at $10GHz$
using the data in Fig. 1. Note that temperature $T$ is the parameter that is varied. This is 
equivalent to a Cole-Cole plot used in dielectric spectroscopy. The plot shows a dominant
relaxation labelled $\beta$ followed by another relaxation labelled $\gamma$ at higher
temperature. The low $T$ $\alpha$ relaxation is not visible on this scale.
The bottom curve represents the data with the $\chi_{c \gamma}^{''}$ contribution subtracted
out leaving a pure $\beta$ relaxation.}
\label{fig3}
\end{figure}%

\begin{figure}
\caption{Spin relaxation rate $1/2 \pi \tau_{\beta} (T)$ {\em vs.} $T$ obtained from the data of Fig. 1. }
\label{fig4}
\end{figure}%

\begin{figure}
\caption{ 
 $\chi_c ^{\prime \prime }(10GHz,T)$ versus $T$ for $Sr_2CuO_3$, $SrCuO_2$
and $Sr_{14}Cu_{24}O_{41}$.}
\label{fig5}
\end{figure}%

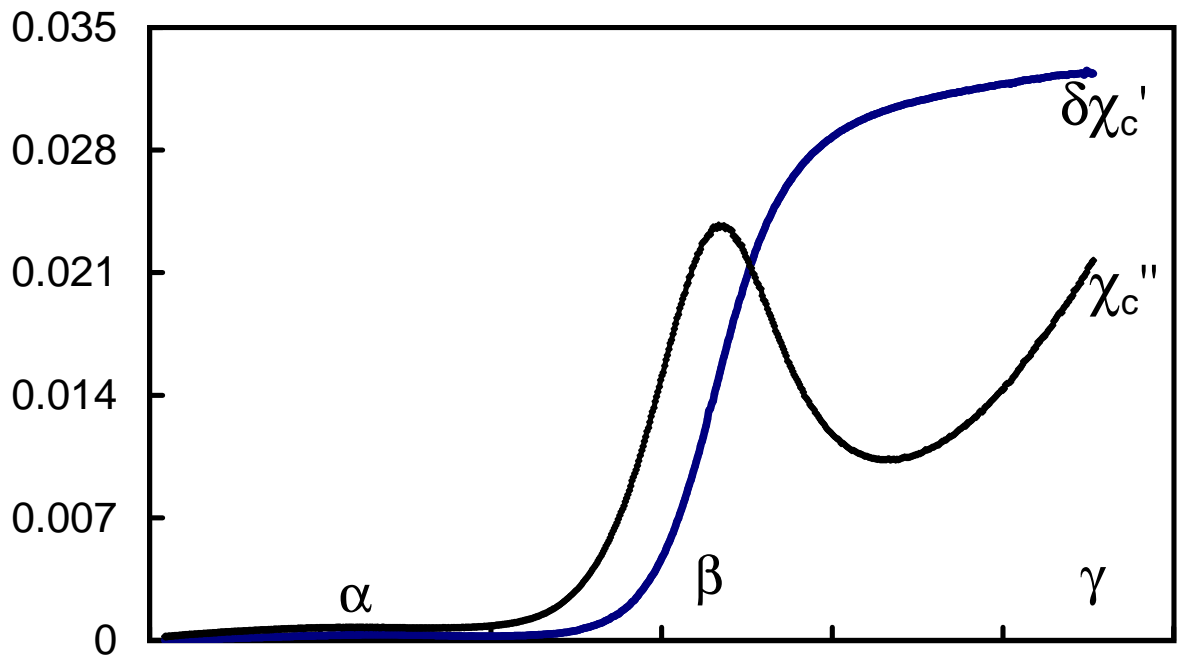
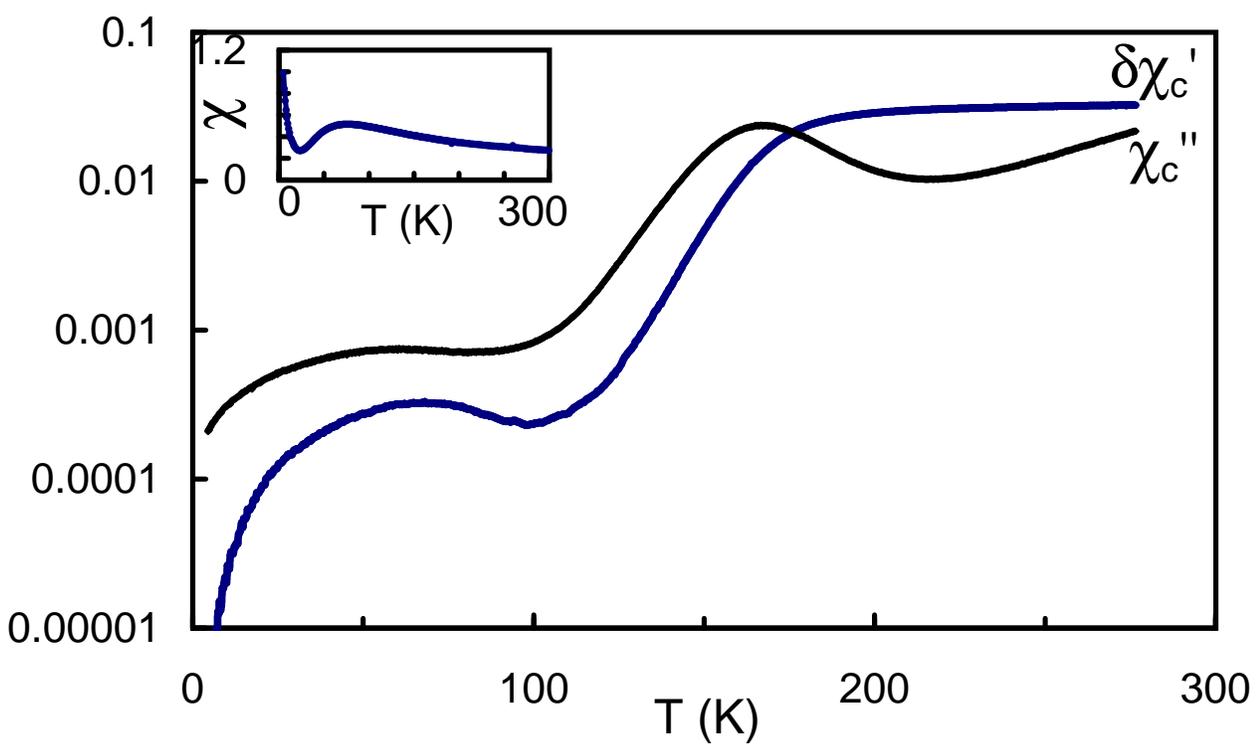
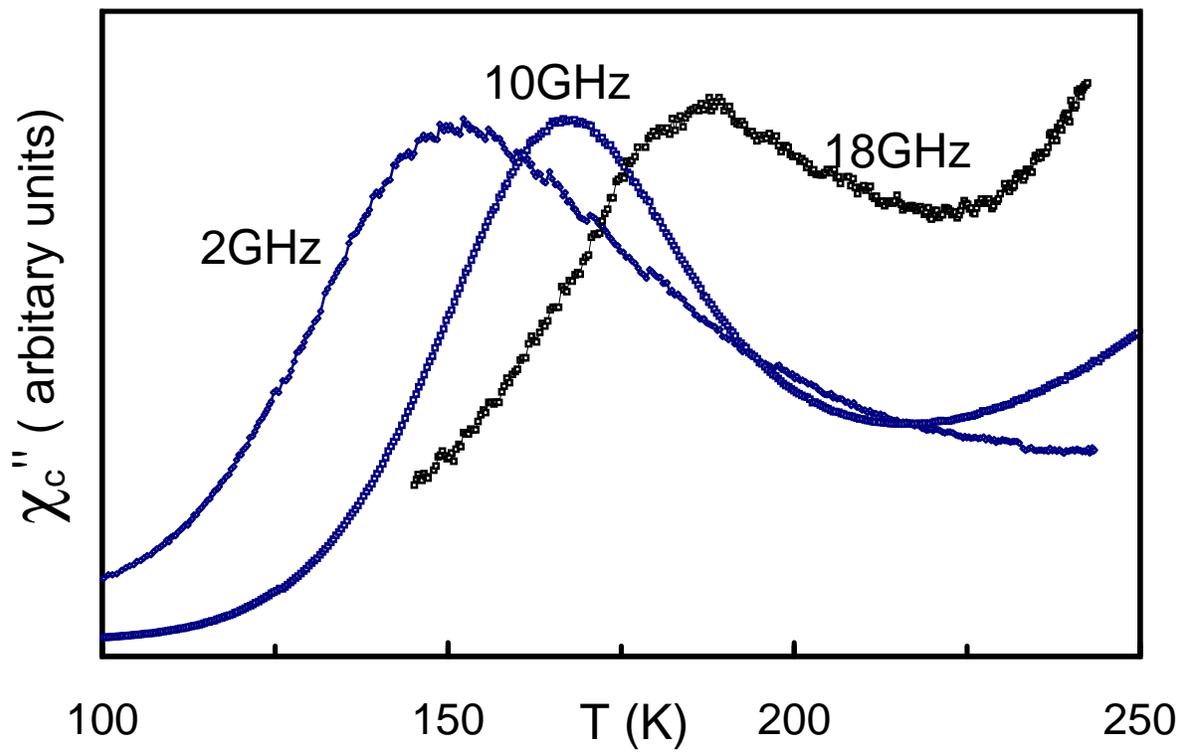
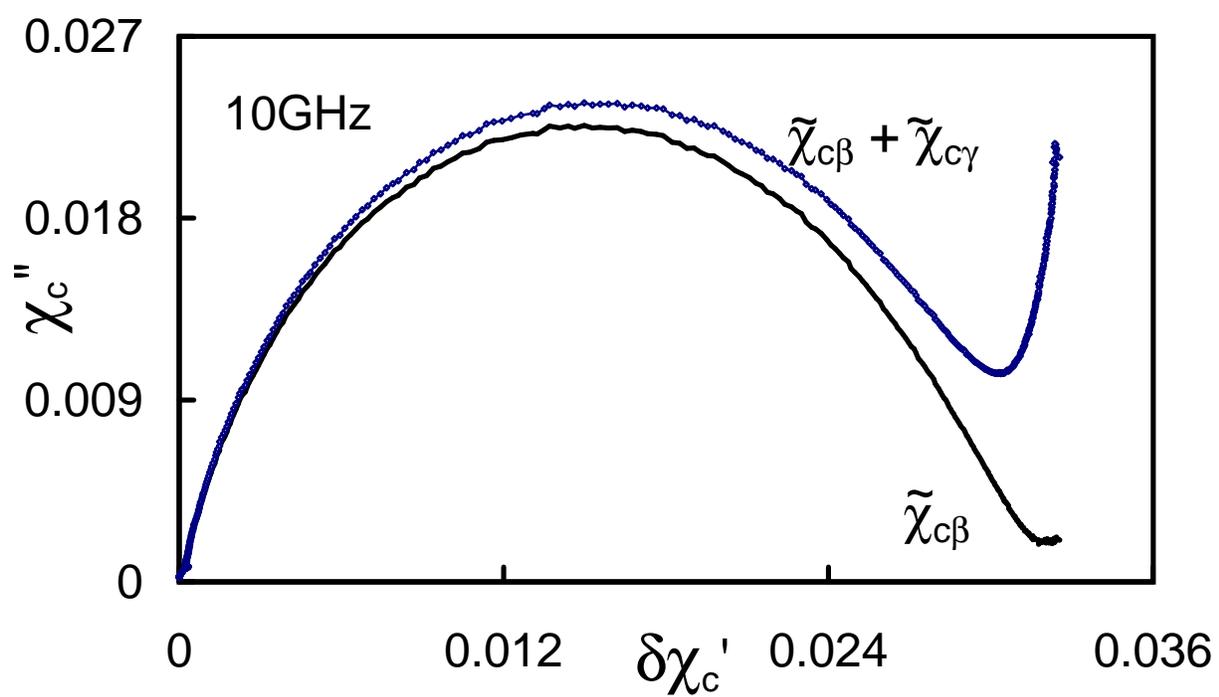
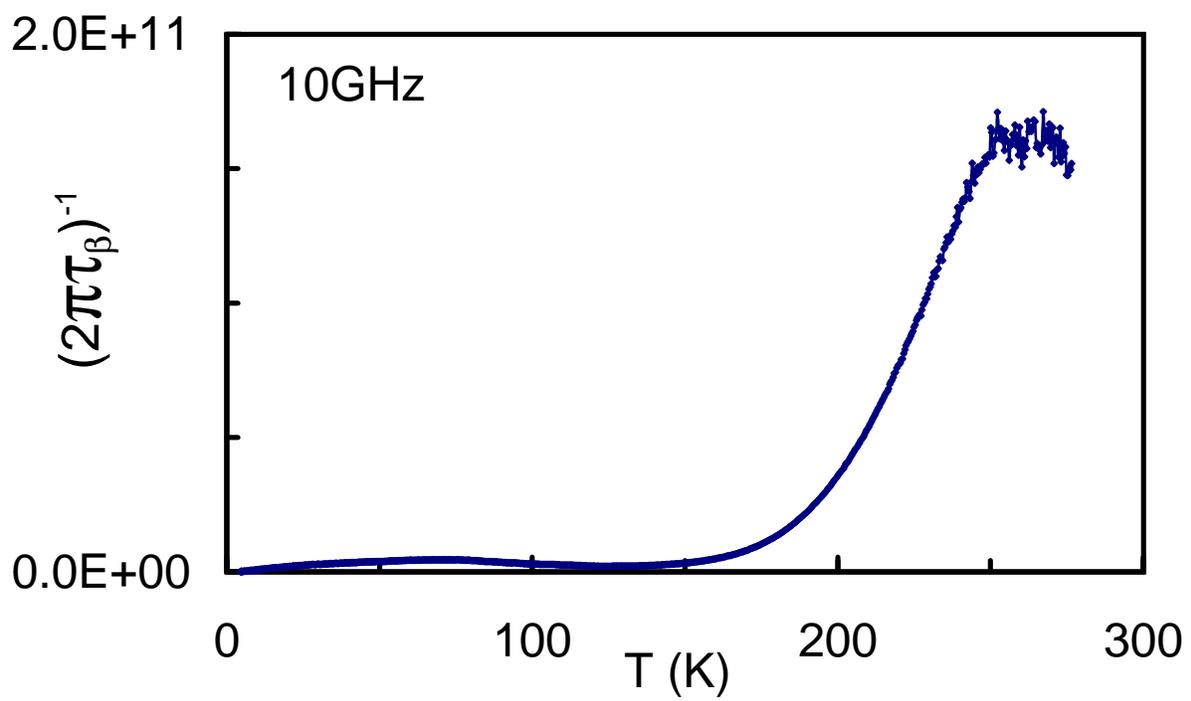
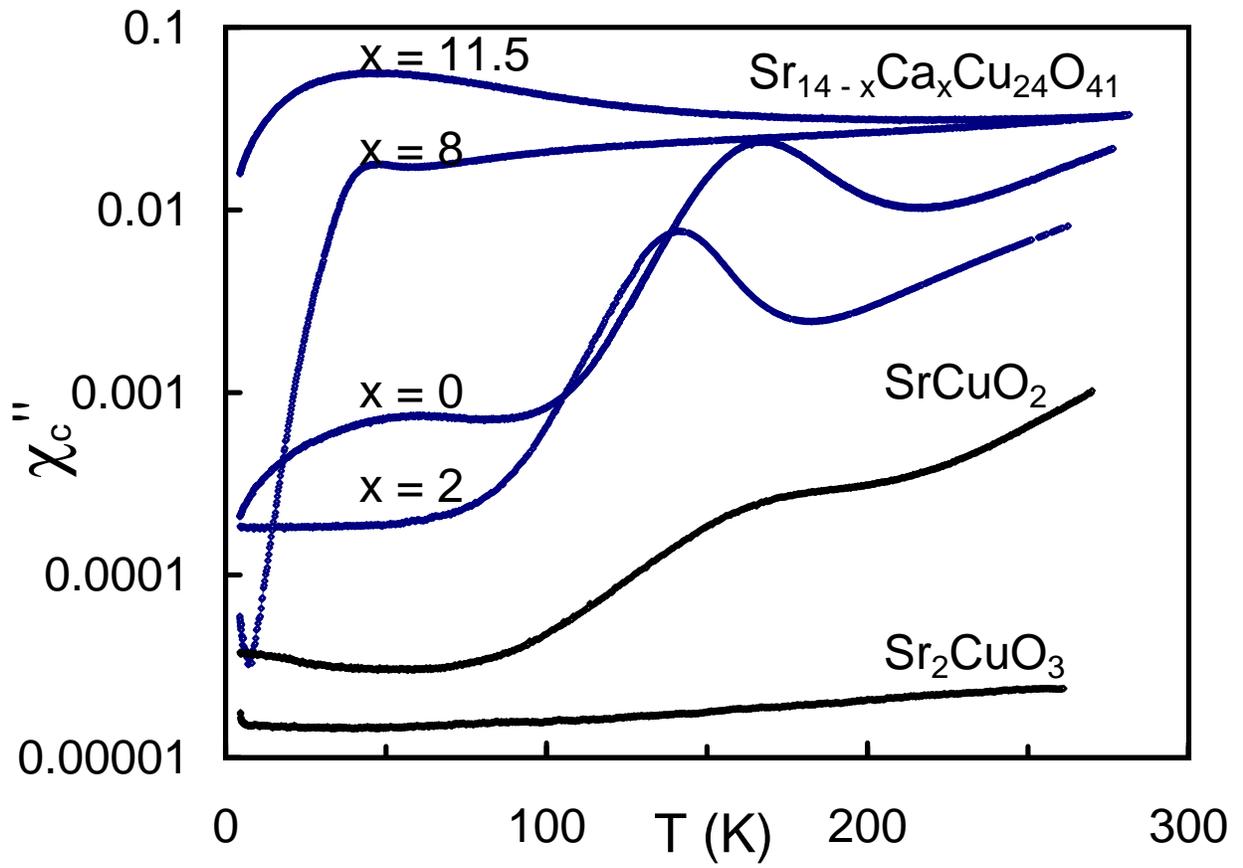
\end{document}